\def\year{2020}\relax
\documentclass[letterpaper]{article} 
\usepackage{aaai20}  
\usepackage{times}  
\usepackage{helvet} 
\usepackage{courier}  
\usepackage[hyphens]{url}  
\usepackage{graphicx} 
\usepackage{graphicx}  
\usepackage{subcaption}

\title{Detecting Troll Behavior via Inverse Reinforcement Learning:\\ A Case Study of Russian Trolls in the 2016 US Election}

\author{Luca Luceri,\textsuperscript{\rm 1,2,3} 
Silvia Giordano,\textsuperscript{\rm 2}
Emilio Ferrara\textsuperscript{\rm 1} \\ %
\textsuperscript{\rm 1} University of Southern California, Information Sciences Institute, Marina del Rey, California \\ \textsuperscript{\rm 2} University of Applied Sciences and Arts of Southern Switzerland (SUPSI), Manno, Switzerland \\ \textsuperscript{\rm 3} University of Bern, Bern, Switzerland\\
luca.luceri@supsi.ch, silvia.giordano@supsi.ch, emiliofe@usc.edu
}

\begin{document}

\maketitle

\begin{abstract}
Since the 2016 US Presidential election,
social media abuse has been eliciting massive concern in the academic community and beyond. 
Preventing and limiting the malicious activity of users, such as trolls and bots, in their manipulation campaigns is of paramount importance for the integrity of democracy, public health, and more. However, the automated detection of troll accounts is an open challenge.
In this work, we propose an approach based on Inverse Reinforcement Learning (IRL) to capture troll behavior and identify troll accounts.
We employ IRL to infer a set of online incentives that may steer user behavior, which in turn highlights behavioral differences between troll and non-troll accounts, enabling their accurate classification.
As a study case, we consider the troll accounts identified by the US Congress during the investigation of Russian meddling in the 2016 US Presidential election.
We report promising results: the IRL-based approach is able to accurately detect troll accounts (AUC=89.1\%).
The differences in the predictive features between the two classes of accounts enables a principled understanding of the distinctive behaviors reflecting the incentives trolls and non-trolls respond to.
\end{abstract}

\section{Introduction}

During the last few years, social media have been facing a huge wave of nefarious activity on their platforms.
For instance, the massive diffusion of digital misinformation and the increasing presence of malicious accounts have been identified as major threats for a healthy online discussion,
other than as primary factors contributing to the manipulation of public opinion on social media \cite{ferrara2015manipulation}.
Examples of social media manipulation can be found in a variety of contexts, ranging from politics to public health \cite{allem2017cigarette,chen2020covid}, e.g.,  diffusion of false news, guerrilla advertising, anti-vaccination movements, and stock market manipulation.
In particular, the risk of mass manipulation of public opinion is creating enormous concerns in the political sphere, where social media abuse has been shown to harm the democracy of online discussions and may affect the sovereignty of the election 
\cite{vosoughi2018spread,stella2018bots,metaxas2012social,del2017mapping,howard2016bots}.

One of the most striking cases, in the political context, regards the Russian interference in the online discussion during the 2016 US Presidential election.
Russia has been accused of using bots (i.e., software controlled accounts) and trolls (i.e., human operators tied with information operation agencies) to spread propaganda and politically biased information.
In this regards, the US Congress disclosed a list of 2,752 (now-deactivated) Twitter accounts that have been identified as tied to the Russia's ``Internet Research Agency'' (IRA).
Russia's troll farm constitutes the first revealed case of human operators assembled to carry a deceptive online interference campaign.
The activity of such accounts has been extensively documented by the research community, highlighting their attempts at sowing division along divergent narrative frames \cite{badawy2019characterizing,gerber2016does,badawy2018falls,dutt2018senator,Badawy2018,zannettou2019disinformation,broniatowski2018weaponized,stewart2018examining,kim2019analysing}.

Most recently, Twitter identified and suspended malicious accounts originating from Russia, Iran, Bangladesh, and Venezuela, suggesting the presence of efforts to manipulate online discourse across countries \cite{gadde2018enabling}. 
Although the attempt from social media providers to purge their platform, malicious accounts persist \cite{im2019still,luceri2019red}, conducting manipulation campaigns over diverse global political events \cite{ferrara2017disinformation}
and multiple consecutive elections \cite{luceri2019evolution}. 
Preventing and limiting the activity of such operators is of fundamental importance for the integrity of voting events.
While researchers have brought to the table different approaches for identifying bot accounts \cite{davis2016botornot,varol2017online,yang2019arming,chavoshi2016debot,subrahmanian2016darpa,chen2018unsupervised,kudugunta2018deep},
the detection of trolls remains an open challenge. Zannettou \textit{et al.}
\cite{zannettou2019let} found that the behavior of state-sponsored trolls varies over time: automatically unveiling their activity is not straightforward and researchers have not yet established a principled solution to the detection of troll accounts.

\subsection*{Contributions of this work}

In this paper, we propose an approach for identifying troll accounts
based on Inverse Reinforcement Learning (IRL).
Our solution only
relies on the flow of users' activity on social media. 
In particular, we aim to consider whether and how users' online activity is affected by the interaction with peers and, more specifically, by the feedback they receive from other users in the social network, e.g., a user might be more motivated to share new content if she receives positive feedback from others.
The rationale is to unveil the incentives driving trolls' activity and utilize such cues for distinguishing them from non-troll accounts (henceforth, simply referred to as \textit{users}). 
The objective of inferring intent and motivation from observed behavior has been extensively studied under the framework of IRL.
Specifically, IRL has the main goal of finding the rewards  behind an agent's observed behavior \cite{piot2016bridging}.
In our scenario, we employ IRL to 
infer the rewards that could have led trolls and users to perform their online activity. We then consider to exploit the 
estimated rewards as features of a supervised learning algorithm aimed at classifying such accounts.
To evaluate the proposed approach, we consider the Twitter accounts identified by the US Congress as tied to the IRA's troll farm.
Results corroborate our intuition and highlight peculiar characteristics steering trolls' online behavior. Our work conveys the following three contributions:

\begin{itemize}
\item We show that an established user behavioral model \cite{wang2016unsupervised} based on users' clickstream (i.e., sequence of online actions) is not suitable for discerning trolls' activity. According to the outcome of the clickstream clustering \cite{wang2016unsupervised}, trolls seem to behave similarly to other users, hence appear indistinguishable from them.

\item We present a novel approach for detecting the activity of trolls on social media. To the best of our knowledge, the presented approach is the first attempt to use IRL for classification purposes and, in particular, to identify malicious accounts. The IRL-based classification approach is able to accurately detect Russian troll accounts with an AUC of 89.1\% based only on the flow of their online activities.
\item The rewards estimated with IRL show significant behavioral differences between trolls and users, suggesting that these two classes of accounts respond to different sets of incentives.  
In particular, Russian trolls differ in their behavior when engaged by other users or when their content is re-shared.
They perform their activity regardless of the feedback they get from non-trolls and merely focus on the spread of the content they generate.
\end{itemize}

\section{Data}
\label{dataset}
To attain a set of  trolls that serves as ground truth, 
we rely on the list of 2,752 Twitter accounts identified as Russian trolls by the US Congress and publicly released\footnote{\textit{Recode's Twitter's list of 2,752 Russian trolls}. See: \url{https://www.recode.net/2017/11/2/16598312/russia-twitter-trump-twitter-deactivated-handle-list}} during the investigation of Russian interference in the 2016 Presidential election.
To recover trolls' tweets, we leverage the dataset collected by the research community \cite{addawood2019linguistic,badawy2019characterizing}. The authors utilized Crimson
Hexagon, a platform that provides paid
datastream access. This allowed the authors to obtain tweets generated by trolls and subsequently deleted after their
suspension 
from Twitter \cite{addawood2019linguistic,badawy2019characterizing}.
The dataset presents 1,148 Russian trolls who posted over 1.2M tweets. 

Our dataset also includes non-troll users'  tweets, which have been collected by the researchers \cite{addawood2019linguistic,badawy2019characterizing} by utilizing a set of keywords (employed also in \cite{bessi2016social}) related to the 2016 US Presidential election  event. Also, to capture users' baseline behavior (not strictly related to the political context), the authors gathered tweets from the same users that do not include the political-based keywords. This collection yielded 12,361,285 tweets produced by 1,166,760 users.

We pre-processed this dataset to filter out accounts engaged with just a few tweets.
In particular, we consider only users and trolls that shared at least $k=10$ posts and were involved in at least $k=10$ other accounts' posts (retweet, reply, or mention).
This allowed us to build, for each account, a time-ordered sequence of tweets (of at least 20 elements) in which the account is involved both actively and passively.
Such filtering resulted in 342 trolls and 1,981 users.
In the next sections, we describe how we aim to analyze the sequence of users' online activity to characterize their behavior and recognize malicious accounts.

\section{User Behavior Analysis}
The focus of this paper is to unveil distinctive behaviors and incentives of trolls with respect to other users to enable their detection.
Notice that our goal is to identify malicious accounts controlled by humans and that, thus, do not rely on automation.
Several studies try to provide an understanding of human behavior through their online activity \cite{lerman2010using,lerman2012using,wang2016unsupervised,das2014effects}.
In \cite{wang2016unsupervised}, the analysis of users' clickstream on social media 
is used to identify common behaviors.
The authors propose an unsupervised method
to categorize behavioral patterns and cluster users accordingly.
More specifically, \textit{clickstream clustering} \cite{wang2016unsupervised} identifies clusters of similar users by partitioning a similarity graph, whose nodes represent users and edges among them are weighted to capture users' activity similarity. 
The authors use a hierarchical clustering approach 
and an iterative feature pruning technique to detect behavioral patterns and build a hierarchy of clusters, where higher-level (resp. lower-level) clusters encode more general (resp. specific) behavioral patterns.
This approach has been proven to outperform existing clustering methods (e.g., K-means) in user behavior analysis. 
It has been shown that such model can help service providers to identify unexpected behavior such as malicious accounts and hostile chatters. 

Therefore, with the objective of characterizing and detecting malicious trolls' activity, we attempt to model the online behavior of social media accounts by leveraging the clickstream clustering approach.
The input of the clustering approach is represented by the sequence of online actions performed by every account. We determine these activities according to the sharing options available on Twitter, i.e., tweet, retweet, reply, and mention.
The output of this approach is a tree hierarchy of behavioral clusters populated by the accounts under investigation.

In our scenario, the clickstream clustering results in 5 disjointed clusters, 
which do not present any hierarchical pattern among them, suggesting that such groups are strongly distinguished.
The clickstream clustering approach also outputs the \textit{Action Patterns}, which characterize the behavior of the accounts in every cluster. 
The Action Patterns provide a statistical description of how accounts within a given cluster are different from accounts outside of the cluster based on their activities.
Basically, the Action Patterns explain how clusters have formed and which user behavior every cluster encompasses. Therefore, by inspecting the output of the clustering approach, we aim to understand whether and how trolls can be distinguished from users.
The 5 clusters (named C1, C2, C3, C4, and C5) differ from each other as follows:

\begin{itemize}
    \item[\textbf{C1}:] composed of 504 accounts (17.0\% trolls). Accounts in this cluster perform \textit{reply} and \textit{retweet} less frequently than the accounts outside the cluster.
    \item[\textbf{C2}:] composed of 162 accounts (15.4\% trolls). Accounts in this cluster \textit{tweet} more frequently and \textit{retweet} less frequently than the accounts outside the cluster.
    \item[\textbf{C3}:] composed of 261 accounts (13.8\% trolls). Accounts in this cluster perform \textit{mention} more frequently than the accounts outside the cluster.
    \item[\textbf{C4}:] composed of 1034 accounts (17.1\% trolls). Accounts in this cluster \textit{retweet} more frequently than the accounts outside the cluster.
    \item[\textbf{C5}:] composed of 362 accounts (5\% trolls). Accounts in this cluster perform \textit{reply} more frequently than the accounts outside the cluster.
    
\end{itemize}

From the above clustering outcome, it appears that trolls behave similarly to other users. In fact, troll accounts are embedded in every of the above clusters 
with a comparable percentage (except from C5), which makes them indistinguishable from users.

This finding motivates us to explore other approaches for unveiling such malicious accounts. Although online activities represent users' behavior in the social network, their analysis takes only into account individual behavior.
In such a way, we do not consider whether and how users are influenced by the interaction with others \cite{anagnostopoulos2008influence,luceri2019analyzing,luceri2017social} and, more specifically, by the feedback they receive from their peer \cite{crandall2008feedback,das2014effects}. 
The effects of social feedback on online activities have been studied in \cite{das2014effects}, where the authors argue how the potential endorsement from other users may trigger or retrain online activities. For example, a user might be more motivated to share new contents if she receives positive feedback.
In a similar way, we aim to understand the driving forces behind the online activities of trolls and users with the purpose of understanding whether behavioral differences can arise between these two classes of accounts. 
The objective of inferring intent and motivation form observed behavior has been extensively studied under the framework of IRL \cite{reddy2018you}. Therefore, in this study we rely on the IRL paradigm, which we detail in the next Section.

\section{Methodology}

In this Section, we first provide an overview of the IRL paradigm. We then discuss how we leverage IRL to characterize the online activity of social media accounts.
We finally present the IRL formulation to assess users' and trolls' incentives (i.e., rewards),
which are subsequently used for the detection purpose.

\subsection{Background}
\label{IRL_background}

In this subsection, we present the background needed to comprehend the proposed IRL-based approach to detect trolls. 
IRL is a machine learning framework that aims at solving the inverse problem of Reinforcement Learning (RL).
Therefore, we first formalize the RL paradigm, which will guide the subsequent presentation of IRL.

\subsubsection{Reinforcement Learning}
\label{rl}
RL is defined as the problem faced by an agent that learns how to behave through trial-and-error interactions with an uncertain environment \cite{kaelbling1996reinforcement}.
The objective of the agent is to achieve a goal by performing a sequence of actions in the environment. The agent learns the actions to take 
according to some rewards or penalties. 

The most used and straightforward way to formulate a RL problem is to frame the environment as a finite Markov Decision Process (MDP), which includes:

\begin{itemize}
    \item A finite set of states $S$, which represents the situation in which the agent finds itself in the environment;
    
    \item A finite set of actions $A$ that the agent can perform in the environment;
    
    \item Transition probabilities $T$ between states, which represent the probability to move from one state to another when performing a certain action;
    
    \item A finite set of scalar rewards $R$ associated with each transition;
    
    \item A discount factor $\gamma$, which determines the importance between short-term and long-term rewards.
\end{itemize}

\begin{figure}[t] 
\centering
  \includegraphics[width=0.95\columnwidth]{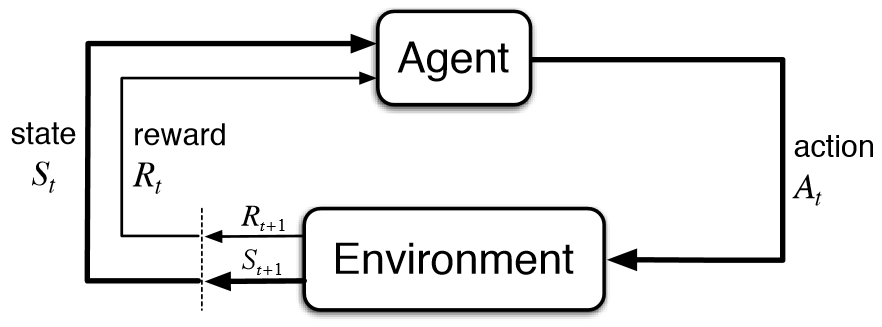}
  \caption[RL schema]{RL schema \cite{sutton2018reinforcement}}
  \label{fig:rl}
\end{figure}

The general framework of a RL problem is displayed in Fig. \ref{fig:rl}.
The agent (e.g., a robot) has to learn how to achieve a goal (e.g., the end point of a maze) from the interaction with the environment (e.g., a maze with rewards and penalties).
The agent and the environment interact at each step $t$ of a sequence of discrete time steps ($t=0,1,2, \dots $) \cite{sutton2018reinforcement}.
The interaction flow follows the schema displayed in Fig. \ref{fig:rl}: At each step $t$, the agent select an action $A_t \in A$ based on its current state $S_t \in S$. The environment responds to such action and presents to the agent a new state $S_{t+1} \in S$ and a reward $R_{t+1} \in R$ associated with the transition $(S_t,A_t)$.
As time steps goes by, a sequence, called \textit{trajectory} $\zeta$, as the following arises: $\zeta= \{(S_0,A_0),(S_1,A_1),(S_2,A_2),\dots\}$.

\subsubsection{Inverse Reinforcement Learning}
\label{irl_def}

While in RL the agent is provided with a reward function, which is used to achieve a given goal, the rationale of IRL is to find the rewards that explains the observed behavior of the agent. 
This objective has two main motivations \cite{ng2000algorithms}.
The first reason is for imitation purpose (i.e., to replicate the agent's behavior), when a reward function is not explicitly determined \cite{ramachandran2007bayesian}.
The second reason is to support behavioral studies, where computational models are needed for understanding human and animal behavior \cite{watkins1989learning,touretzky1997operant,das2014effects}.

Formalizing, the objective of IRL is to estimate a reward function that could have led to agent's actions.
More specifically, and relying on the MDP framework described above, the observed behavior of the agent is expressed as the history of its trajectory $\zeta$.
The goal is to uncover the hidden reward function that maps each state-action pair $(s,a)$ to a real value $r_{s,a} \in R$.
Therefore, IRL aims to find a function $g$ such that: $g(S,A)=R$.
In many applications, however, the size of the state space does not allow to compute a
reward function for every state-action pair.
Therefore, current IRL techniques leverage features that approximate the state-action space
to capture the structure of the reward function 
\cite{ng2000algorithms,abbeel2004apprenticeship,ratliff2006maximum,levine2010feature}.

The \textit{Maximum Entropy IRL} \cite{ziebart2008maximum} represents the most popular solution to solve IRL problems.
This approach maps a set of features $f$ to the reward $R$ as a weighted linear combination of the feature values:
\begin{equation}
\label{max_irl}
    g(f)=g(f,\theta)=\theta^\mathrm{T} f=R.
\end{equation}
Most recent approaches propose non-linear models to represent the reward structure \cite{wulfmeier2015maximum,choi2013bayesian,levine2011nonlinear}. In particular, Wulfmeier \textit{et al.} \cite{wulfmeier2015maximum} introduce the usage of Convolutional Neural Networks in the context of IRL in an approacch called \textit{Maximum Entropy Deep IRL}. This approach has showed comparable performance with respect to Maximum Entropy IRL and is considered particularly suited for life-long learning scenarios \cite{wulfmeier2015maximum}.

\subsection{Twitter MDP}
We here propose to model the social media Twitter with a MDP.
More specifically, we represent Twitter as an environment constituted of multiple agents (i.e., Twitter accounts) that interact with each other to achieve a certain goal (e.g., to spread content, increase their popularity, or influence other people).
The interaction between every agent and the environment abides by the RL schema described in Section \textit{Reinforcement Learning}.
The agent performs a certain action (e.g., share a content) and receive a feedback from the environment (e.g., the content is re-shared by other accounts).
In such environment, we consider that the following four actions can be performed by the agents:
\begin{itemize}
    \item Generate original content. We refer to this action as \textit{active tweet} (\textit{tw}).
    \item Re-share content generated by others. We refer to this action as \textit{active retweet} (\textit{rt}).
    \item Interact with other users by means of reply or mention. We refer to this action as \textit{active reply} (\textit{rp}).
    \item Keep silent. We refer to this action as \textit{active nothing} (\textit{nt}).
\end{itemize}

The Twitter environment, in turn,
responds to such actions and presents to the agent a new state. We represent the set of states with the three following feedback the environment can provide to the agent:
\begin{itemize}
    \item Re-share agent's tweet. We refer to this state as \textit{passive retweet} (\textit{RT}). 
     \item Interact with the agent by means of reply or mention. We refer to this state as \textit{passive reply} (\textit{RP}). 
     \item Do not engage with the agent. We refer to this state as \textit{passive nothing} (\textit{NT}). 
\end{itemize}

We leverage the history of the accounts on Twitter to analyze their interaction with the environment.
To this aim, we sort in chronological order every online activity in which the account is involved, either actively (actions) and passively (states). Therefore, at each step (i.e., an element of the ordered sequence of the account's online activities) the agent can be in one of the above mentioned states and perform only one action. 
We consider the agent to execute action \textit{nt} (active nothing) if it does not react to the environment feedback, e.g., it does not perform \textit{tw}, \textit{rt}, and \textit{rp}.  
Similarly, we represent with \textit{NT} the case in which the environment does not react to agent's action. 

\subsection{Users and Trolls IRL}
\label{usr_trl_irl}

As we mentioned in Section \textit{Inverse Reinforcement Learning}, the state-action space is usually represented by a set of features $f$. 
Similarly to \cite{rhinehart2018first}, we build $f$ by concatenating variables related to the set of states and actions.
Specifically, we define $f$ as the possible combinations of the following variables:
$($\textit{RT}, \textit{RP}, \textit{tw}, \textit{rt}, \textit{rp}$)$, where the first two features represent the state space, while the last three features indicate the action space. Each feature is a binary variable, which assumes value 1 based on the state in which the agent is and the action it performs (e.g., if the agent is in the state \textit{RT} the first feature assumes value 1, while the second feature equals to 0). 
The state \textit{NT} is represented by setting the first two features to zero, while the last three features equal to zero indicates the action \textit{nt}.
As an example, the tuple $(0,0,0,0,1)$ indicates that the agent performed an active reply (\textit{rp}) while it was in state $NT$.
Given that at each step the agent can be in only one state and performs only one action there exists 
12 possible combinations of state-action pairs. As an example, the pair (\textit{RT},\textit{tw}) describes the scenario where the environment returns a retweet and the agent generates a new tweet.
Therefore, $f$ is a $5\times12$ matrix, where 5 is the number of features and 12 is the number of state-action pair possible combinations.   
In turn, the reward function is a $1\times12$ vector, where each element represents the scalar reward of a certain state-action pair.

Based on the ordered list of states and actions, we build for each account a trajectory $\zeta$ composed of the feature representation of state-action pairs, 
as discussed above. 
Such trajectory represents the observed behavior of the agent. 
We then utilize IRL approaches to estimate the reward function that drove agent's behavior.
The rationale is to investigate whether users and trolls share the same rewards or, oppositely, are guided by different incentives.
To this end, we employ two IRL approaches:
Maximum Entropy IRL \cite{ziebart2008maximum} and its non linear variation, i.e., Maximum Entropy Deep IRL \cite{wulfmeier2015maximum}.
We deploy such IRL approaches by leveraging \cite{alger16}.
It should be noticed that every account is modeled independently from the others. This means that a unique IRL model has been developed for each account.
Each model takes as inputs the feature matrix $f$, the trajectory $\zeta$, and the transition probabilities $T$. As discussed in Section \textit{Reinforcement Learning}, the latter
represents the probability of transition from state $s \in S$ to state $s' \in S$ after performing action $a \in A$.
We compute $T$ by observing the occurrences of $(s,a,s')$ triplets in the account's trajectory. 
Both the IRL approaches use these inputs to compute the reward function $R$, which determines the scalar reward for each state-action pair. We do not present in details the methodology behind the employed IRL solutions as it is out of the scope of this paper. The interested reader can refer to \cite{ziebart2008maximum,wulfmeier2015maximum}.
In the next subsection, we explain how we employ the estimated rewards for detecting trolls activity.

\subsection{Trolls Detection}
\begin{figure}
    \centering
    \includegraphics[width=.95\columnwidth]{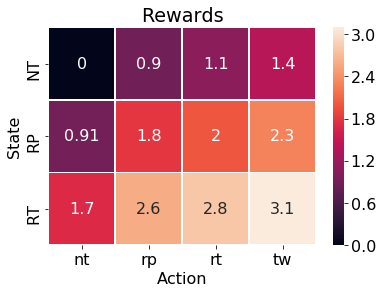}
    \caption{Example of estimated rewards: the x-axis displays the set of actions, the y-axis the set of states, and each value represents the estimated reward for the state-action pair.}
    \label{fig:s}
\end{figure}
For every account, 
IRL approaches output 12 scalar values related to the state-action combination described above.
As an example, Fig. \ref{fig:s} depicts the estimated rewards related to the online activity of a generic user.
Each value represents the reward that could have led the account to perform a given action in a certain state.
A high reward in the generic state-action pair $(s,a)$ indicates that the user is very motivated to perform action $a$ in state $s$.
The opposite holds for low reward values.

Our approach aims to understand whether some differences can be noticed in the motivation and incentives between trolls and users.
We hypothesize that reward values might highlight distinctive behavioral characteristics between troll and user accounts.
Therefore, we propose to utilize the 12 reward values to discriminate these two classes of accounts.
More specifically, we utilize the estimated rewards as features of a supervised learning algorithm aimed at detecting troll accounts. We frame this problem as a classification task, where the two classes are \textit{troll} (positive class) and \textit{user} (negative class). We test several off-the-shelf machine learning algorithms to perform such classification, whose results are discussed in the next Section.

\section{Results}
In this Section, we first discuss the results of the IRL-based classification approach for detecting troll accounts on social media. Then, we analyze the incentives that users and trolls respond to and we highlight the behavioral differences between these two classes of accounts. 

\subsection{Account Classification}
The dataset under investigation presents imbalanced classes, with the negative class (more than) 5 times larger than the positive one (1,981 vs. 342 accounts).
To solve the data imbalanced issue, we employ the undersampling technique \cite{liu2008exploratory,drummond2003c4}, which uses only a random set of the larger class in the classification task. Therefore, we split the negative samples in five parts and we repeat our evaluation every time with a different subset of negative samples.
Then, to train and test our model, we utilize a 10-fold cross validation preserving the percentage of samples for each class.

\begin{table}[t!]
\centering\small
\caption{IRL approaches performance comparison:
AUC of different classifiers fed with the rewards computed with the Maximum Entropy IRL \cite{ziebart2008maximum} and the Maximum Entropy Deep IRL \cite{wulfmeier2015maximum} approaches.} 
\begin{tabular}{@{}lcc@{}}
\hline

& \textbf{Max. Entropy} & \textbf{Max. Entropy Deep} \\
\hline
K-Neighbors & 83.2\%    & 82.4\%  \\  
SVC & 74.2\%  &  85.4 \% \\  
Gaussian Process & 83.8\%     & \textbf{85.6}\%\\ 
Decision Tree & 82.7\%     & 74.1\% \\  
MLP & 84.4\%   & 79.8\% \\  
AdaBoost& \textbf{89.1\%}     & 83.3\% \\
Random Forest & 86.7\% &  81.3\% \\
Naive Bayes & 79.3\%     & 78.7\% \\
\hline
\end{tabular}
\label{IRL_results}
\end{table}

We evaluate the classification (troll vs. user) results obtained by using the two different IRL approaches to estimate the rewards. As we previously mentioned,
we test the Maximum Entropy IRL \cite{ziebart2008maximum} and the Maximum Entropy Deep IRL \cite{wulfmeier2015maximum}.
We utilize the rewards inferred by each of these approaches to feed a supervised learning classification algorithm.
In particular, we compare several off-the-shelf machine learning approaches.
Performance, in terms of Area Under the Curve (AUC), of the different classifiers and IRL approaches are depicted in Table \ref{IRL_results}.
In most of the cases, the Maximum Entropy IRL approach achieves better accuracy with respect to the Maximum Entropy Deep IRL solution.
Thus, we continue our analysis by leveraging the rewards estimated with Maximum Entropy IRL.
Also, we rely on AdaBoost as it outperforms the other supervised learning algorithms. We tune the AdaBoost classifier by performing a grid search of its hyper-parameters, whose best configuration involves 500
weak estimators and a learning rate of 0.05.
The combination of Maximum Entropy IRL with AdaBoost achieves an AUC of 89.1\%.
To further investigate the classification performance of such a solution, we consider the trade-off between sensitivity (i.e., the proportion of correctly identified trolls) and specificity (i.e., the proportion of correctly identified users) by measuring the True Positive Rate (TPR) and the True Negative Rate (TNR), respectively.
Our approach achieves an average TPR
of 79.5\% and an average TNR of 83.2\%.
These results suggest that the proposed solution 
builds a conservative classification model that is mainly targeted at the minimization of false positives (i.e., users classified as trolls). 
This represents a favorable asset of the proposed methodology, as the inaccurate classification of organic users might elicit both ethical concerns and an overhead for social media providers engaged in purging their platforms from malicious entities.

\begin{figure}[t!]
    \centering
    \includegraphics[width=.95\columnwidth]{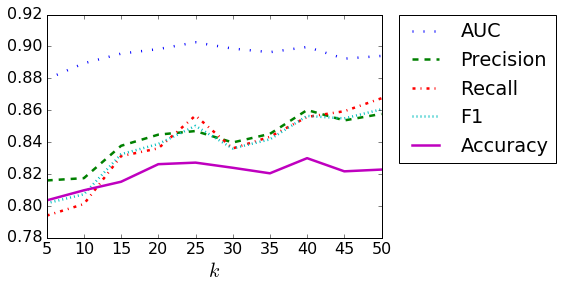}
    \caption{Classification performance at varying $k$.}
    \label{perf}
\end{figure}

We further evaluate the performance of our approach by considering classification metrics such as \textit{accuracy}, \textit{precision}, \textit{recall} (or TPR), F1 and AUC at varying $k$.
As we mentioned in Section \textit{Data}, we filtered out accounts that were involved in less than $k=10$ active and passive online activities.
Here, we aim to examine the impact of $k$ on the classification performance.
In Figure \ref{perf}, we display the classification performance of the proposed solution (Maximum Entropy IRL with AdaBoost) in terms of the classification metrics mentioned above.
As we could expect performance improves as $k$ increases as the Maximum Entropy IRL approach can rely on more information per each user. 
It is also noticeable that, when $k=10$ (i.e, the scenario considered up to this point), precision is higher than recall. This confirms the conservative scheme of the IRL-based model towards the minimization of false positives, as highlighted before in the comparison between TPR and TNR. Interestingly, the gap between precision and recall diminishes with increasing $k$ and their values nearly converge when $k>10$. This finding shows that the proposed model enhances its ability of identifying trolls when more information about accounts' online activity is available.
Overall, our detection approach, even with a small amount of information, achieves prominent performance in every classification metric. This supports our intuition of using IRL to analyze online behavior and to identify malicious troll accounts accordingly.

\begin{figure}
    \centering
    \includegraphics[width=.95\columnwidth]{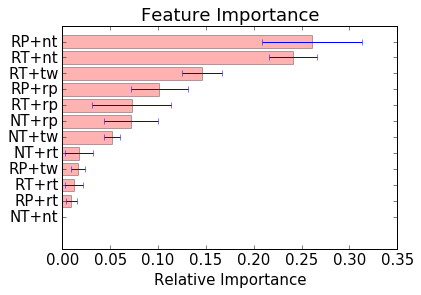}
    \caption{AdaBoost feature importance.}
    \label{feature_importance}
\end{figure}
\begin{figure*}[t!]
    \centering
    \includegraphics[width=0.8\textwidth]{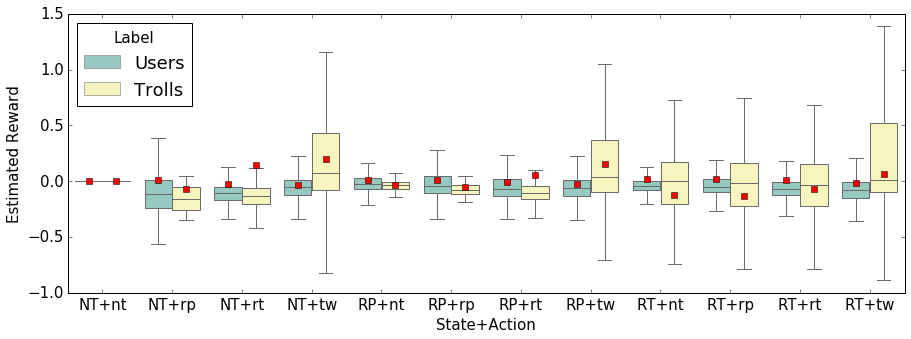}
    \caption{Distribution of the estimated rewards by Maximum Entropy IRL.}
    \label{fig:rewards}
\end{figure*}

To investigate the most distinguishing characteristics between trolls' and users' behavior, we observe the feature importance of the AdaBoost algorithm.
In Fig. \ref{feature_importance}, the relative importance of each feature (i.e., reward) is displayed. Interestingly, the most important features are related to the case
when the account is in the state \textit{RT} (i.e., its content have been re-shared by other accounts) or \textit{RP} (i.e., it is engaged by other users).
To shed light on the significance of such cues, in the next subsection, we provide a statistical assessment of the estimated rewards of troll and user accounts.

\subsection{Rewards Analysis}
To better understand the difference in trolls' and users' intent, we compare their estimated rewards. Figure \ref{fig:rewards} displays the distribution of the rewards assessed with Maximum Entropy IRL and highlights some differences in the behavior of the two classes of accounts.\footnote{It should be noticed that the displayed rewards have been normalized by using the StandardScaler technique (i.e., each features is standardized by removing the mean and scaling to unit variance).}
This discrepancy is also demonstrated by a two-sample Kolmogorov-Smirnov test, which in turn unveils significant differences ($p < 0.01$) between every pair of distributions, except for the state-action pair (\textit{NT},\textit{nt}), (\textit{NT},\textit{rt}), and
(\textit{RP},\textit{rt}). By looking at Fig. \ref{feature_importance}, we can notice how these three rewards are in the 5 least important features for the AdaBoost classifier. 
From Fig. \ref{fig:rewards}, it is possible to notice a flat distribution for the pair (\textit{NT},\textit{nt}). This is due to the fact that we do not observe any instance of such scenario in users' trajectory, given that we can only leverage data that show users' involvement (either active or passive) in the tweets.

Furthermore, in Fig. \ref{fig:rewards},
it can be appreciated how, on average, trolls are more motivated (i.e., higher reward) to share a new original tweet (action \textit{tw}) with respect to users regardless of their state, which suggest that trolls merely focus on spreading their content, independently from others' feedback.
Also, the most noticeable discrepancy is appreciable in the actions performed when the account is in the \textit{RT} state, which is in line with the feature importance results displayed in Fig. \ref{feature_importance}.

\begin{figure}[t!]
    \centering
    \includegraphics[width=.75\columnwidth]{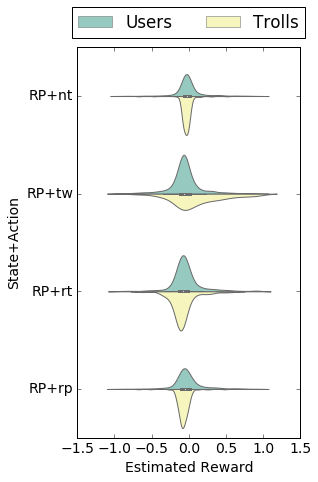}
    \caption{Violin plot of the estimated rewards in the \textit{RP} state.}
    \label{rp_state}
\end{figure}

The latter also showed that the \textit{RP} state is relevant in the discrimination between trolls and users. To take a closer look at the features related to this state, in Fig. \ref{rp_state}, we depict the violin plot distribution of the estimated rewards related to \textit{RP} state.
It can be observed how the distributions of troll and user accounts particularly differ in the case of action \textit{nt} and \textit{rp}, which are the first and forth most relevant features of the AdaBoost classifier. 

\begin{figure}
    \centering
    \includegraphics[width=\columnwidth]{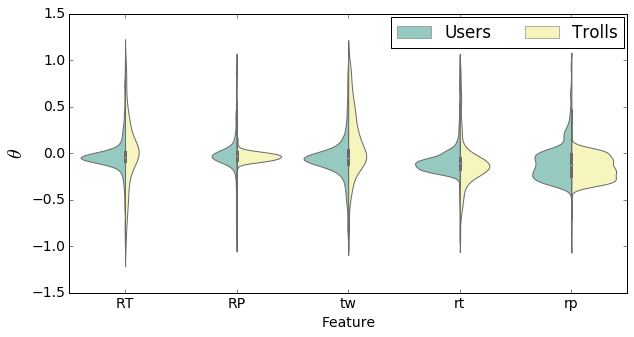}
    \caption{Distribution of weights $\theta$ for each feature in $f$.}
    \label{fig:feature_weight}
\end{figure}

\begin{figure}[t!]
    \centering
    \includegraphics[width=.9\columnwidth]{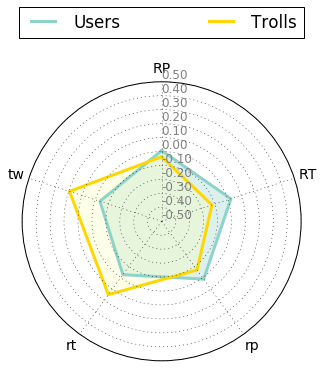}
    \caption{Comparison of $\theta$ average values of trolls and users.}
    \label{fig:spider}
\end{figure}

Moreover, being a linear model, the Maximum entropy IRL approach allows to disentangle the reward of each state-action pair. This is done by solving Eq. \ref{max_irl}, given that $f$ is known and $R$ has been estimated with Maximum entropy IRL.
This operation allows us to recover the
weights $\theta$ that represent a measure of importance for each feature in $f$.
The weights distribution for each class of accounts is displayed in Fig. \ref{fig:feature_weight}. The violin plot distribution further highlights behavioral differences between trolls and users, which are proved to be significant for every feature $(p<0.01)$ by means of a two-sample Kolmogorov-Smirnov test. 
Additionally, Fig. \ref{fig:feature_weight} confirms the differences of behavior between trolls and users when they are in state \textit{RT} and \textit{RP}, consistently with our previous findings.
Also, the action that shows the most noticeable discrepancy between the two groups of account is related to the active tweet (\textit{tw}). 

Finally, in Fig. \ref{fig:spider}, we depict the mean value of such distributions. The purpose is to compare the relevance of each feature between troll and user accounts.
On average, trolls appear to be more motivated than users in generating new content (\textit{tw}) and re-sharing others' post (\textit{rt}). This finding is in line with trolls' purpose of spreading (false) pieces of information and harm the online conversation.
On the other hand, users appear to be more rewarded than trolls when their content is re-shared by other accounts (\textit{RT}). This, representing a form of social endorsement \cite{metaxas2015retweets,stella2018bots}, might suggest that users are more concerned about others' esteem with respect to trolls.

\section{Related Work}

\subsection{Russian Trolls}
Trolls, along with bots, are the principle actors in manipulation and misinformation campaigns on social media.
The 2016 US Presidential
election raised awareness of such issue and, in particular, spotlighted 
the
activity of malicious operators
allegedly funded by the IRA. 

Badawy \textit{et al.} \cite{badawy2019characterizing} showed that Russian trolls aimed to harm the political conversation and create distrust in the political system. These users operated with the objective of influencing conversations about political issues and creating discord among different groups \cite{gerber2016does,popken2018twitter,stewart2018examining}.
In \cite{stewart2018examining}, it is shown how trolls acted to accentuate disagreement and sow division along divergent frames, as further confirmed by Dutt \textit{et al.} \cite{dutt2018senator} in relation to Russian ads on Facebook.
In \cite{Badawy2018}, the authors studied the effects of the manipulation campaign by analyzing the accounts that endorsed trolls activity on Twitter. They found that conservative leaning users re-shared trolls content 30 times more than liberal ones.
Zannettou \textit{et al.}
\cite{zannettou2019disinformation} compared trolls behavior with other (random) Twitter accounts recognizing differences in the content they spread, in the evolution of their accounts, and in the strategy adopted to increase their impact. 

Furthermore,
the nefarious activity of Russian trolls has not only been recognized in the political context. As an example, in \cite{broniatowski2018weaponized}, the authors study the attempt of such operators to undermine the public health in the vaccine debate. Trolls further evidenced their strategy of promoting discord, in this scenario across pro- and anti-vaccination arguments.

\subsection{Malicious Accounts Detection}

The detection of malicious accounts represents a pivotal asset in the fight against the abuse of social media.
Bots and trolls are known to be used in coordinated campaigns~\cite{monsted2017evidence,stella2018bots,zannettou2019let,addawood2019linguistic,im2019still,kim2019analysing}, whose detection is hard~\cite{ferrara2016detection,varol2017early}.
The research community offered
multiple approaches for identifying automated accounts \cite{davis2016botornot,varol2017online,yang2019arming,chavoshi2016debot,subrahmanian2016darpa,chen2018unsupervised,kudugunta2018deep}.
Differently from bots, the automated detection of trolls is still an open challenge. 
Zannettou \textit{et al.} \cite{zannettou2019let} investigate state-sponsored trolls on Twitter and Reddit. In particular, the authors focus on Russian and Iranian trolls trying to characterize their activity and strategies.
They found that the behavior of such trolls is not
consistent over time, thus, their automated identification is not a straightforward task.

Concurrently with our undertaking, two approaches for unveiling the activity of Russian trolls on Twitter have been proposed \cite{addawood2019linguistic,im2019still}.
To identify trolls attempt to manipulate the public opinion during the 2016 US election, Addawood \textit{et al.} \cite{addawood2019linguistic} identified 49 linguistic markers of deception and measured their use by troll accounts. They show that such deceptive language cues can help to accurately identify trolls. In \cite{im2019still}, the authors proposed a detection approach that relies on users' metadata, activity (e.g., number of shared links, retweets, mentions, etc.), and linguistic features to identify active trolls on Twitter.

Our work is similar to the above mentioned efforts in the objective of unveiling troll accounts. However,
differently from \cite{addawood2019linguistic,im2019still}, we do not aim at spotting differences in trolls' language or metadata.
We do not 
exploit such signals to identify trolls, but we focus on uncovering hidden behavioral differences between troll and non-troll accounts.
As we do not leverage either social network-dependent metadata or language features, our approach can potentially be generalized on different social media and to state-sponsored trolls originating from different countries.
In fact, our solution only
relies on the sequence of users' activity on online platforms to capture the incentives the two classes of accounts respond to.

A sequence analysis approach has also been proposed in \cite{kim2019analysing} for classifying the social roles of trolls.
The authors exploit the temporal and semantic similarity in the sequence of shared content (i.e., tweet) to classify trolls into the categories (e.g., left-leaning, right-leaning, news feed) defined in \cite{boatwright2018troll}. We share with this work \cite{kim2019analysing} the idea of discerning classes of accounts based on their activity traces. While Kim \textit{et al.} \cite{kim2019analysing} consider text and time as features to categorize the troll population into subgroups, our undertaking estimates the motivation behind the sequence of online activities to discern troll and non-troll accounts, without considering content and temporal differences in the tweet sequences.

\section{Discussion}
In this paper, we showed that Russian trolls, although not relying on automated activity, do not act and behave as organic users. Their distinct behavior enabled their accurate identification and a principled understanding of the incentives behind their online activities.
However, given the rationale of our detection approach, trolls might alter their regular activity and implement evasion strategies to avoid detection. 
For instance, trolls might decide to mutate their behavior by operating similarly to other users. 
This, however, could potentially affect trolls' main purpose of harming online discussion, which, in turn, may limit the effectiveness of their malicious activity.
Such discussion, and related adversarial analysis, opens the door to diverse research directions, which we plan to explore in future work.

Another limitation of the presented approach is related to the size of the dataset used in our analysis, which, in turn, pertains to a single political event (i.e., 2016 US Presidential election). We recognize that a larger and different set of data would be needed to validate our approach across sets of trolls with diverse origins and purposes. However, this represents a challenging objective for two reasons, at the minimum.
First, trolls associated with state-agencies are needed as ground truth to verify the accuracy of our approach. Therefore, the identity of such accounts (e.g., in terms of username) should be disclosed by social media providers or other entities (e.g., the US Congress in the case of the 2016 US Presidential election).
Second, corresponding non-troll accounts (same country, context, topics of discussion) are needed as negative samples to learn a model and enable the accounts classification.
As an example, the set of data recently released by Twitter \cite{gadde2018enabling} includes only information about the activity of malicious users. However, the flow of online activities performed by organic users and the interactions among every kind of account represent the necessary ingredients to fuel our approach. 
These challenges represent fertile ground for future work aimed to provide an exhaustive validation of the presented solution.

Nevertheless, in this paper, we have presented a first attempt to identify state-sponsored (Russian) trolls by uncovering 
the incentives driving their behavior.
The IRL-based model provides a generalizable approach, which is agnostic to the OSN platform and to the nature of the troll accounts, which both represent a breakthrough with respect to existing solutions for the detection of troll accounts. Also, and not secondarily, our solution provides an explainable model, which is in line with the emerging paradigm of eXplainable Artificial Intelligence (XAI) \cite{gunning2017explainable,pedreschi2019meaningful}. Indeed, by leveraging IRL, our approach is capable of unveiling and discerning the most peculiar behavioral characteristics (i.e., the features of the classification algorithm) between troll and user accounts that enable their accurate classification.
Given that such a methodology is also agnostic to the specific phenomenon of online trolling, IRL may represent a useful addition to the toolbox of user behavior analysis.
This asset, combined with the generalizable nature of the proposed solution, allows us to envision this detection system to be used in various domains (e.g., politics, health, etc.) and 
along different dimensions, not only related to the manipulation of public opinion in the political context, but also, and more generally, in the fight against every form of online harassment and abuse in social media (e.g., cyberbullying).

\section{Conclusion}
Social media manipulation is an issue of paramount importance, potentially harming the integrity of the online discourse and, in the political context, hampering the democratic process.
Bots and trolls represent the most recognized categories of malicious accounts acting in such efforts. 
While the detection of bots has established solutions, the automated identification of troll accounts has been proven to be a challenging (yet unsolved) task.

In this paper, we presented an approach to detect trolls' activity on social media.
The proposed solution is based on Inverse Reinforcement Learning (IRL) and only relies on the flow of online activity. IRL allowed us to characterize the behavior of social media actors by inferring the reward structure (or motivational affordances) behind their actions. These cues are further employed as inputs of a supervised learning algorithm aimed at classifying troll and non-troll accounts.
Considering the set of trolls identified by the US Congress as tied with the Russia's IRA, we show that our approach accurately separates trolls from other users by leveraging and exploiting the diverse motivations between the two classes of accounts.
The IRL model also allowed us to recognize and unveil the most distinctive behavior that differentiate troll from non-troll accounts. For example, Russian trolls and users differ in their behavior when they are engaged by other users or their content is re-shared.
Also, troll accounts appear to perform their sharing activity irrespective of others' feedback and simply focus on the spread of the content they generate.

This paper represents a first step in the direction of understanding, characterizing, and detecting trolls, which is a critical challenge in the race towards  healthy online ecosystems.
In our future endeavors, we aim to validate the proposed approach with data from other known state-sponsored trolls. Also, we aspire to extend the classification task to a diverse set of malicious entities motivated by different incentives.

\smallskip\textbf{ Acknowledgments}.
Emilio Ferrara gratefully acknowledges support by the Air Force Office of Scientific Research (award no. FA9550-17-1-0327) and DARPA (grant no. D16AP00115). Luca Luceri and Silvia Giordano are supported by the Swiss National Science Foundation via the CHIST-ERA project UPRISE IoT.

\bibliography{sample-base}
\bibliographystyle{aaai}

\end{document}